\documentclass[twocolumn,prl,preprintnumbers,amsmath,amssymb]{revtex4}
\usepackage{color}
\usepackage{graphicx}
\usepackage{bm}

\begin{document}
 
\title{Nonlinear dynamics and millikelvin cavity-cooling of levitated nanoparticles}

\author{ P. Z. G. Fonseca, E. B. Aranas, J. Millen, T. S. Monteiro, and P. F. Barker}
 \affiliation{Department of Physics and Astronomy, University College London, Gower Street, London WC1E 6BT, United Kingdom}

\begin{abstract}
Optomechanical systems explore and exploit the coupling between light and the mechanical motion of matter. A nonlinear coupling offers access to rich new physics, in both the quantum and classical regimes. We investigate a dynamic, as opposed to the usually studied static, nonlinear optomechanical system, comprising of a nanosphere levitated and cooled in a hybrid electro-optical trap. An optical cavity offers readout of both linear-in-position and quadratic-in-position (nonlinear) light-matter coupling, whilst simultaneously cooling the nanosphere to millikelvin temperatures for indefinite periods of time in high vacuum. We observe cooling of the linear and non-linear motion, leading to a $10^5$ fold reduction in phonon number $n_p$, attaining final occupancies of $n_p = 100-1000$. This work puts cavity cooling of a levitated object to the quantum ground-state firmly within reach.


\end{abstract}

\maketitle

Cavity optomechanics, the cooling and coherent manipulation of mechanical oscillators using optical cavities, has undergone rapid progress in recent years \cite{AMKreview}, with many experimental milestones realized. These include cooling to the quantum level \cite{Lehnert11,Painter11_2}, optomechanically induced transparency (OMIT) \cite{OMIT}, and the transduction \cite{Painter11,Cleland13,Lehnert14} and squeezing \cite{Painter13} of light. These important processes are due to a {\em linear} light-matter interaction; linear in both the position of the oscillator $\hat{x}$ and the amplitude of the optical field $\hat{a}$.

\begin{figure}[t]
{\includegraphics[width=3.in]{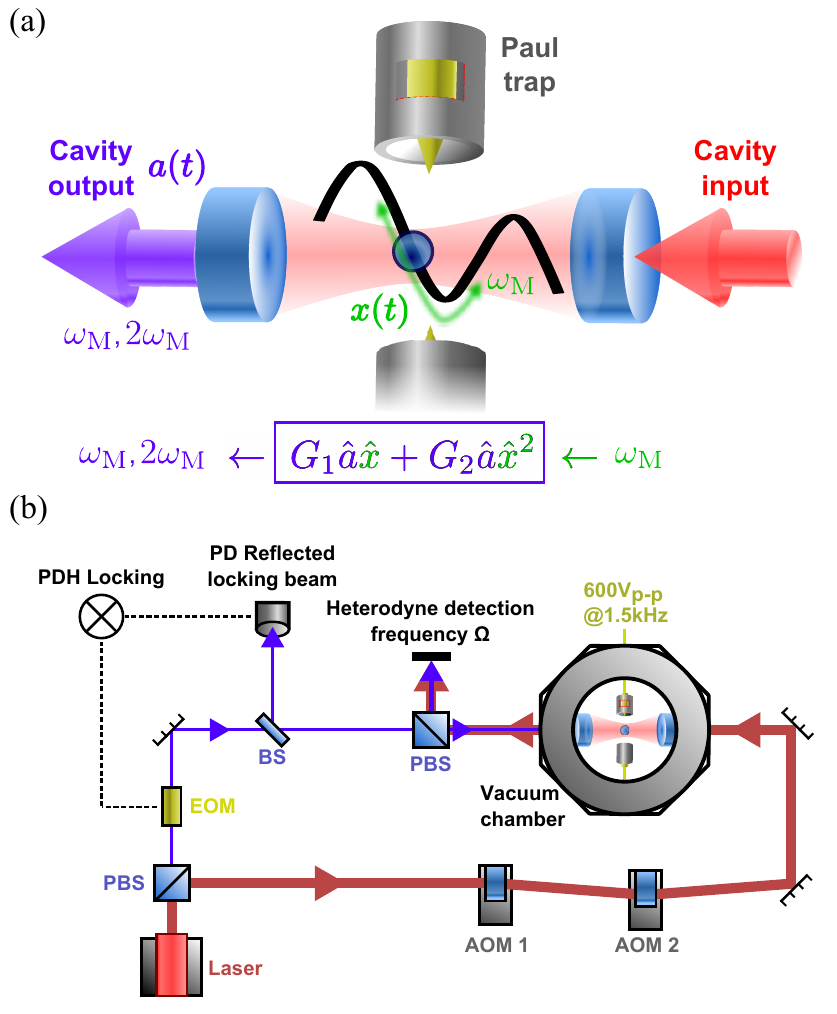}}
\caption{ Concept and experiment. {\bf (a)} A particle trapped in an optical well of the hybrid trap
experiences an optomechanical light-matter coupling with linear and quadratic components of variable
strengths. The motion of the particle modulates the optical cavity output.
{\bf (b)} Light that is resonant with the optical cavity (used to stabilize the cavity) is coincident on a detector with light that is red detuned (for cooling and trapping), leading to a beat signal with heterodyne frequency $\Omega/(2\pi)$.}
\label{Fig1} 
\end{figure}

{\em Nonlinear} optomechanical interactions open up a new range of applications which are so far largely unexplored. In principle, they allow quantum nondemolition (QND) measurements of energy and thus the possibility of monitoring quantum jumps in a macroscopic system \cite{Thompson2008,AMKreview}. They also offer the prospect of observing phonon quantum shot noise \cite{Clerk10}, nonlinear OMIT \cite{Nonlintheory1,Nonlintheory2}, and the preparation of macroscopic nonclassical states \cite{jacobs09}. To achieve a nonlinear interaction one can use optical means, which require strong single-photon coupling to the mechanical system \cite{Nonlintheory1,Nonlintheory2} but are a considerable experimental challenge. Nonlinearities can also arise from spatial, mechanical effects, by engineering, for example, a light-matter interaction of the form $(\hat{a}+ \hat{a}^\dagger)(G_1 {\hat x} + G_2 {\hat x}^2)$. Previous studies investigated the {\em static} shift in the cavity resonant frequency \cite{Thompson2008,Doolin14,Harris15} or the quadratic optical spring effect \cite{Harris15} arising from a nonlinear coupling. However, these studies identified the problem of a residual linear $G_1$ contribution to the coupling. Not only can $G_1$ allow unwanted back-action, but a large $G_1^2$ contribution (e.g. \cite{Doolin14}) can mask the signatures of true nonlinear $G_2$ coupling.

In this work, we study a nanosphere levitated in a hybrid system formed from a Paul trap and an optical cavity \cite{PRL2015} as shown in fig 1. The output of the cavity is used to access the linear and nonlinear dynamics of the particle. We are able to tune the $G_1:G_2$ ratio to reach $G_2 \gg G_1$, isolating the true nonlinear dynamics. Further, due to the {\em dynamic} nature of this experiment, we are able to observe the cooling, in time, of the nonlinear contribution to motion. To our knowledge, such damping of $G_2$ has not been
previously observed, and $G_2$ effects have not been previously detected in any levitated system. We stress that these nonlinear dynamical effects are unrelated to variations in mechanical oscillation frequency arising when the particle samples anharmonicities 
in the potential \cite{Gieseler13}, although we note these are also observable in our data.

Due to an enhancement in the light-matter coupling as compared to our previous work \cite{PRL2015}, which exposes the
$G_2$ coupling, there is greatly enhanced (linear) optomechanical cavity cooling. This enables us to show for the first time permanent and stable optical cavity trapping of a levitated nanoparticle at pressures limited by our current equipment $\sim 10^{-6}$\,mbar, where we infer millikelvin temperatures. Finally we identify a previously unobserved shift of the Paul trap secular frequencies due to the optical cavity, which we show gives valuable information on the system, such as the nanoparticle charge and mean number of photons in the cavity.

The optomechanical cooling of levitated nanoparticles to the quantum regime has been the subject of several recent theoretical  \cite{Oreos,Chang10,Barker10,Pender12,Monteiro13} and experimental \cite{Li11,Gieseler12,Kiesel13,Asenbaum13,PRL2015} studies, due to their isolation from environmental decoherence. While there has been success cooling nanoparticles with active feedback \cite{Li11,Gieseler12}, passive cavity cooling has been hindered by particle loss processes which prevented optical trapping below a few mBar \cite{Monteiro13,Kiesel13,Millen14}. This ``bottleneck'' was overcome in \cite{PRL2015} via the use of a hybrid electro-optical trap, and the subsequent improvements presented in this paper allow cooling to temperatures three orders of magnitude lower than previously reported \cite{Kiesel13,Asenbaum13, PRL2015}.

\begin{figure}[ht]
{\includegraphics[width=3.in]{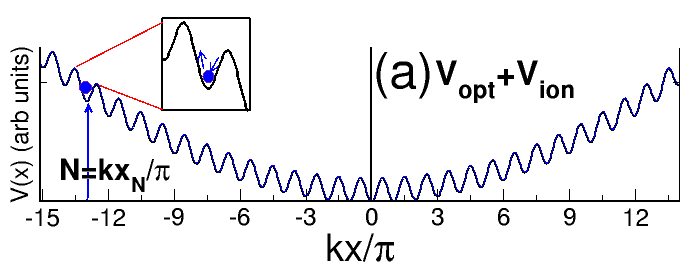}}
{\includegraphics[width=3in]{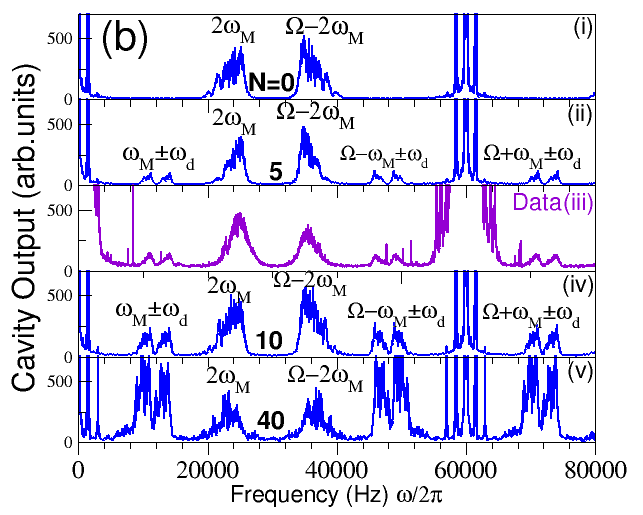}}
\caption{\label{Fig2} 
{\bf (a)} A plot of the hybrid trap potential combining the Paul trap potential
with the standing wave potential of the cavity mode. The cooling rate and the ratio of the $G_1$ and $G_2$ contributions depends strongly on which
optical well $\textrm{N}$ the particle becomes trapped in.
{\bf (b)} Full stochastic simulations (i,ii,iv,v) of the heterodyne power spectra for capture in low wells
 $\mathrm{N}=0-40$  and comparison with an experimental spectrum (iii). All spectra show the strong beat frequency component $\Omega =$ 60 kHz which is the detuning between the on-resonance locking beam and the red detuned cooling beam. The mechanical motion can be observed as sidebands around this peak at $\Omega \pm \omega_\mathrm{M}$ due to $G_1$ coupling and at $\Omega \pm 2\omega_\mathrm{M}$ due to $G_2$ coupling.  There are also peaks at $ \omega_\mathrm{M}$ and $ 2\omega_\mathrm{M}$ due to direct modulation in cavity transmission of the particle. The simulations correspond to a pressure of
 $10^{-2}$ mBar and particle charge of $Q=2$.  For all spectra, the nonlinear contribution is evidenced by dominant strong modulation at $\approx  2\omega_\mathrm{M}$. Comparison of the experimental spectrum with the simulations indicates a well
 capture of $0 < \mathrm{N} < 10$.}
\end{figure}

The hybrid trap consisting of an optical cavity field overlapped by a Paul trap, as illustrated in fig.~\ref{Fig1}, is described as follows. The Paul trap potential for a nanosphere is approximated by
${V(x,y,z,t)= \frac{1}{2}m\omega_\mathrm{T}^2 \left(x^2+y^2 -2z^2 \right) \sin (\omega_\mathrm{d} t)}$,
where $m$ is the mass of the nanosphere and ${\omega_\mathrm{T}^2= \frac{2 Q V_0}{m r_0^2}}$, with $Q$ the charge on the nanosphere, $\omega_d =2\pi \times 1500$\,Hz the drive frequency, $V{_0} = 300$\,V the amplitude of the applied voltage and $r_0= 1$\,mm a parameter setting the scale of the trap potential. This potential is overlapped with the optical cavity potential given by ${V_{\mathrm{opt}}(x,y,z)=\hbar A |\bar{\alpha}|^2 \cos^2{kx} \ e^{-2(y^2+z^2)/w^2}}$
where $|\bar{\alpha}|^2 \equiv \textrm{n}$ is the mean intra-cavity
photon number and the coupling strength  ${A = \frac{3V_\mathrm{s}}{2V_\mathrm{m}}\frac{\epsilon_\mathrm{r}-1}{\epsilon_\mathrm{r}+2} \omega_\mathrm{l}}$
depends on the sphere volume $V_\mathrm{s}$, the mode volume $V_\mathrm{m}=\pi w^2L$ (with $w = 60\,\mu$m the waist of the cavity field and $L = 13$\,mm the length of the cavity) and the laser frequency $\omega_\mathrm{l}$. This combined potential is illustrated in fig.~\ref{Fig2}(a), and shows a particle trapped at optical well $\textrm{N}$. We set the coordinate origin to $x=x_{0}$, where $kx_{0}=\textrm{N}\pi$, with $\lambda = 2\pi/k = 1064$\,nm.

\begin{figure}[t]
{\includegraphics[width=3.in]{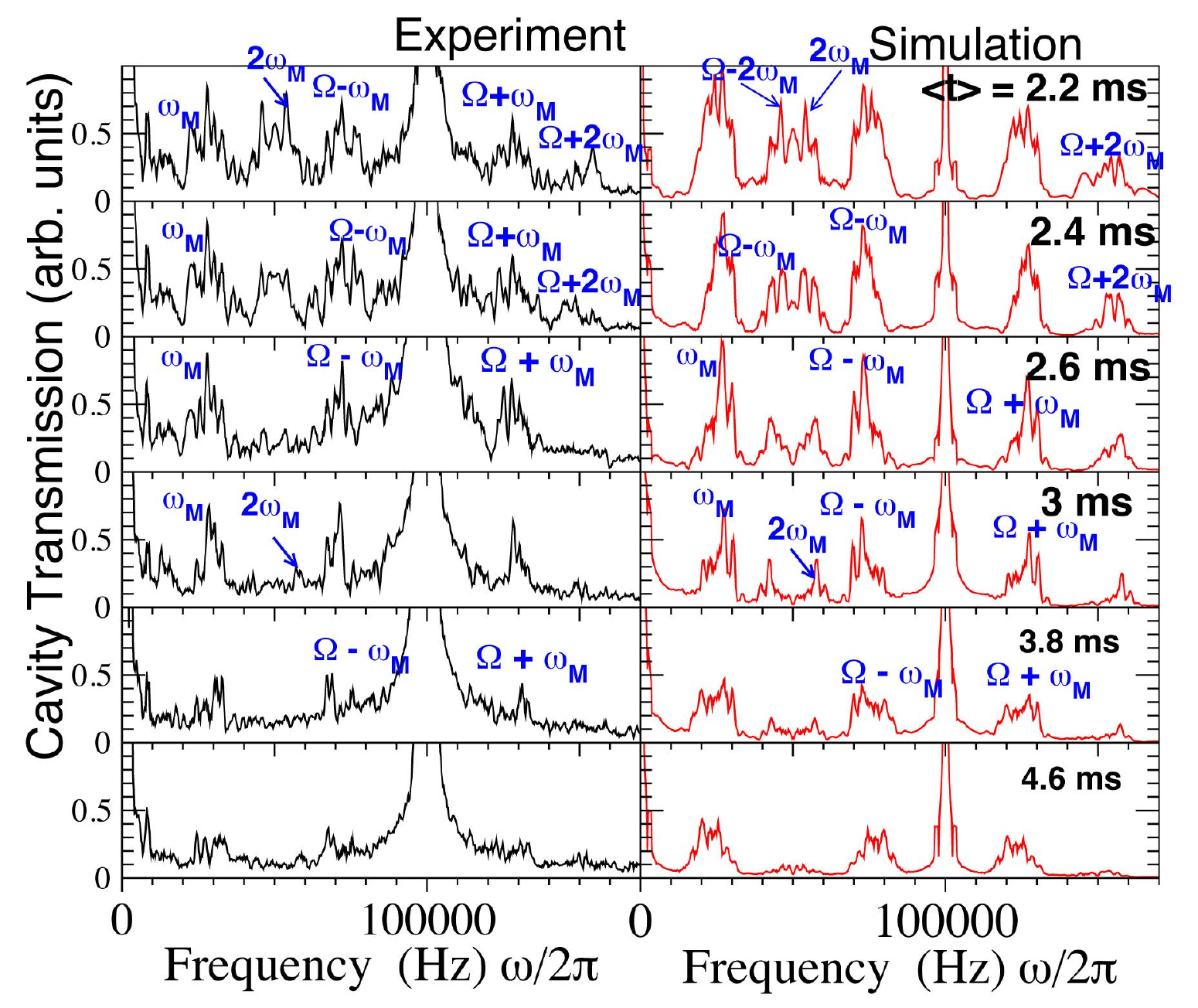}}
\caption{\label{Fig3}Experimental data and simulations of heterodyne power spectra as the particle cools after being captured in well $\textrm{N} \sim 450$, with $\Omega = 100$\,kHz and $Q=1$. Both linear and quadratic modulations by the particle are observed. The nonlinear coupling/frequency-doubled sideband at $2\omega_M$ are visible only in the first few milliseconds while the linear sidebands are damped more slowly. The optomechanical cooling rate for this series is $\Gamma_{\mathrm{opt}} \approx 0.4$ kHz at a pressure $3 \times 10^{-4}$ mBar.}
\end{figure}

The evolution of the axial displacement  $x$ and optical field, $a$, in response to the motion of the particle in the two potentials is given by

\begin{flalign}
&\ddot {x}=\mathcal{W}|a(t)|^2 \sin(2kx)\mathcal{F}(y,z)-\gamma_{\textrm{M}}\dot{x}-\omega^2_{\textrm{T}}x\sin (\omega_{\textrm{d}} t)+ \zeta_x(t)\nonumber\\
&\dot {a}= i\Delta a -i\mathcal{E} +i A a \cos^2(kx)\mathcal{F}(y,z)-\frac{\kappa}{2} a +\eta(t),
\end{flalign}

where $\mathcal{F}(y,z)$ is the envelope of the optical field, ${\mathcal{W}=-\frac{\hbar k A}{m}}$, $\eta(t),\zeta_x(t)$ are stochastic terms, and $\gamma_{\textrm{M}}$ is the damping due to background gas. The $\zeta_x(t)$ terms arise
from the background bath of gas at room temperature with
 ${\langle \zeta_x(t')\zeta_x(t) \rangle \simeq 2\gamma_{\textrm{M}}\frac{k_BT}{m} \delta(t-t')}$,  where $T_B=300$\,K and $\eta(t)$ is the photon shot noise. For simplicity, here we analyse only the axial motion, but a full 3D set of coupled stochastic nonlinear equations were solved in our simulations, with full details given in \cite{SuppInfo}. 

For dynamics about the trapped well co-ordinate we let $x(t)\to x(t)+x_\textrm{0}$. Additionally, we are only interested in the fluctuations in the optical field, $\delta a(t)$, around the static cavity field $\bar{\alpha}$ such that $a(t)\to\bar{\alpha}+ \delta a(t)$. The equations of motion are now given by



 \begin{eqnarray}\label{Xlin}
m\ddot{x}  &\simeq& -m\omega_{\textrm{M}}^2 [x- \epsilon x^2] - \hbar (\delta{a}+\delta{a}^{*})(G_1+ 2 x G_2)  -\gamma_{\textrm{M}}{\dot{x}}\nonumber\\
\dot{(\delta a)} &\simeq&  i\Delta^{x_0} {\delta a} - i (G_1 x +G_2 x^2) -\frac{\kappa}{2} {\delta a},
\end{eqnarray}

where the light-cavity detuning is determined by particle position as ${\Delta^{x_0}=\Delta +  A\cos^2{kx_0}}$, with $\Delta$ the detuning when no particle is present,  
$\mathcal{F}(y,z) \approx 1$, ${m\omega_{\textrm{M}}^2 =  2\hbar k^2 A|\bar{\alpha}|^2 \cos(2kx_0)\nonumber}$ and ${\epsilon=k \tan{(2kx_0)}}$. The position dependent linear and nonlinear couplings are given by ${G_1 = k A \bar{\alpha} \sin(2kx_0)}$ and ${G_2 = k^2 A \bar{\alpha} \cos(2kx_0)}$.

The effect of the oscillating Paul trap field is to force a periodic excursion of the equilibrium point $x_0$:

\begin{equation}
 x_0(t) \approx -\frac{\omega^2_{\textrm{T}}}{2 k \omega_{\textrm{}}^2} 2\pi N \sin(\omega_{\textrm{d}} t),
\label{excursion}
\end{equation}

 with a period which is slow compared with the mechanical oscillations since $\omega_{\textrm{M}}\gg \omega_{\textrm{d}}$.
The amplitude of this oscillation depends on $\textrm{N}$ and this excursion is essential for effective cooling, with a rate:
\begin{equation}
\Gamma_{\textrm{opt}} \simeq G_1^2 \kappa\left[S(\omega_{\textrm{M}})-S(-\omega_{\textrm{M}})\right]
\label{cooling}
\end{equation}
\noindent where $S(\omega)=\displaystyle \left[(\Delta^{x_0}-\omega)^2 +\frac{\kappa^2}{4}\right]^{-1}$.
Thus if $x_0=0$, then $G_1=0$ and there is no cooling, but $G_2$ is maximal. 

A schematic diagram of the hybrid electro-optical trap is shown in fig.~\ref{Fig1}(a). It consists of an optical cavity (finesse $F \simeq 40-55,000$) integrated within a Paul trap inside a vacuum chamber. The Paul trap is formed by two electrodes that are perpendicular to the cavity axis. Silica nanospheres of radius 209 nm (typically with $Q \simeq 1-3$ elementary charges) can be stably trapped in the absence of the optical field.  Nanospheres are introduced into the hybrid trap by initially placing them on an oscillating piezo-disk speaker. Light from a solid state 1064 nm laser is split into a weak and a strong beam formed by a 90:10 beamsplitter, as illustrated in fig.~\ref{Fig1}(b). The weaker beam is used to keep the laser locked to the cavity, via the Pound-Drever-Hall method while the stronger beam is used for trapping and cavity cooling. Its frequency can be shifted with respect to the cavity resonance by using two cascaded acousto-optic modulators.

We trap a particle typically at a pressure of 0.1\,mBar which becomes permanently localised on a cavity anti-node
 at approximately 0.01\,mBar. The particle stays permanently trapped as the pressure is reduced to the current limit of our chamber at $\approx 10^{-6}$\,mBar.  The mechanical frequency of the particle 
($\omega_\mathrm{M}\simeq 2\pi \times 10-40$ kHz in our experiments) can be observed from the heterodyne power spectrum of the recorded time series after the particle is trapped. As the transmitted cooling light from the cavity (red detuned by $\Delta \sim 100$\,kHz) is heterodyned with the on-resonance weak beam reflected from the cavity a strong beat frequency at $\Omega = \Delta$ is observed. The mechanical motion can be observed as heterodyne sidebands around this peak in the spectrum at $\Omega \pm \omega_\mathrm{M}$ due to $G_1$ coupling and at $\Omega \pm 2\omega_\mathrm{M}$ due to $G_2$ coupling.  There are also peaks at $ \omega_\mathrm{M}$ and $ 2\omega_\mathrm{M}$ due to direct modulation of cavity transmission by the particle. 

These features are clearly illustrated in fig.~\ref{Fig2}(b) which contains simulations of the heterodyne spectrum for a nanoparticle captured near to the centre of the optical potential.  For these low N the nonlinear contribution evidenced by strong modulation at $\approx  2\omega_\mathrm{M}$ dominates. Experimental data corresponding to these simulations is shown in fig.~\ref{Fig2}(c) for comparison. Here the splitting of the mechanical frequency and the sidebands by the Paul trap drive can be easily observed as well as the broad single peak at $\approx  2\omega_\mathrm{M}$.

Figure~\ref{Fig3} shows experimental and simulated heterodyne spectra taken at short time periods after a particle is captured in well  $N\sim450$. These illustrate how the heterodyne spectrum changes in time as the particle is cooled at a pressure of $ 3 \times 10^{-4}$ mBar. At this pressure the data shows modest cooling which allows us to follow it in time.   The graphs show the evolution of the heterodyne spectra averaged over a 2.4 ms period separated in time by 0.2 ms. Note that the spectral features are broad due to the short period of time over which the power spectrum is recorded.  There is good agreement between the experiment and theory with both showing the $2\omega_M$ nonlinear coupling/frequency-doubled sidebands which can only be detected in the first few milliseconds as they are rapidly cooled. Cooling of the linear sidebands at  $\omega_M$ occurs more slowly over a $\sim 10$ ms timescale. We obtain a cooling rate of $\Gamma_{\mathrm{opt}} \approx 0.4$ kHz for $Q=1$ which at this pressure cools by factor of $500$ in energy at steady state.

\begin{figure}[t!]
 \includegraphics[width=1.7in]{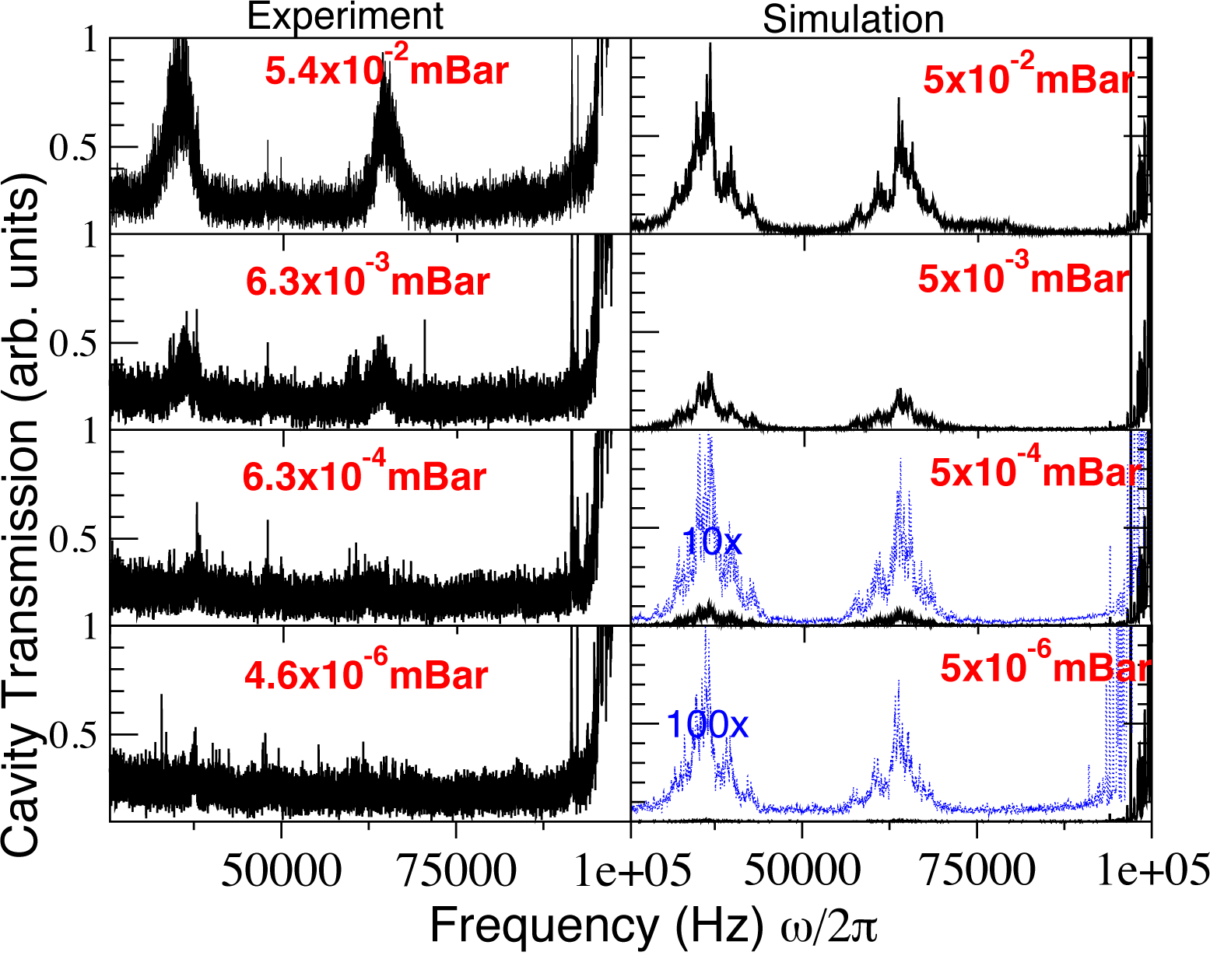}
 \includegraphics[width=1.6in]{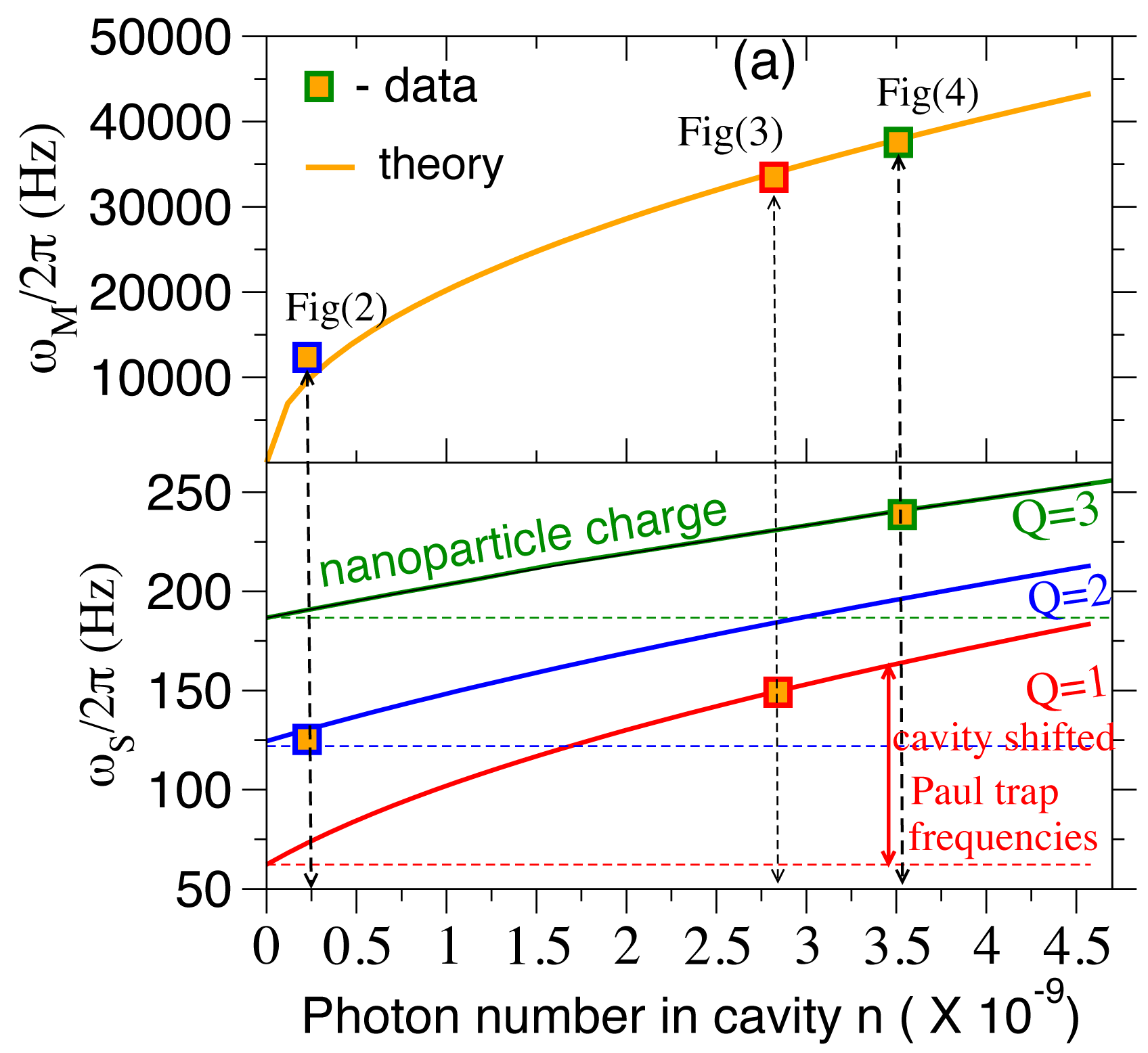}
\caption{{\bf (a)}  Steady-state data and simulations corresponding to a $Q=3$ nanosphere
captured in a high well ($\textrm{N} \simeq 350$) with $\Gamma_{\mathrm{opt}}\simeq 1-2$kHz.  At pressures lower than $\sim 10^{-5}-10^{-6}$ mBar it is no longer possible to read 
the particle's motion from the heterodyne spectra but we infer minimum temperatures of $\sim 1-3$mK  noting
$\textrm{mK}\equiv 270$ quanta above the ground state for $\omega_\textrm{M}/2\pi = 40$ kHz. {\bf (b)} Plots of the mechanical frequency, $\omega_\textrm{M}$, and secular frequency, $\omega_s$, as a function of photon number, $\textrm{n}$, for different particle charge $Q$. The three points  are derived from the data sets presented in figs.~\ref{Fig2}, \ref{Fig3}, \ref{Fig4}(a). These graphs show the novel $\textrm{n}$-dependent shift of the secular frequencies of the Paul trap due to the cavity field. }
{\label{Fig4}} 
\end{figure}

The characteristic signature of high $\textrm{N}$ capture is the gradual
suppression of the  $\omega_M+\omega_d$ peak, resulting in a single dominant peak, which may be asymmetric due to
unresolved sideband structure from smaller, higher order contributions (see \cite{SuppInfo}).
The high cooling rate at high $\textrm{N}$ yields low temperatures in our experiments when the pressure is reduced to the current limit of our apparatus. Such strong cooling is evidenced in the steady state heterodyne spectra shown in fig.~\ref{Fig4}(a). The graphs show the change in the power spectrum for pressures from approximately 10$^{-2}$ to 10$^{-6}$ mbar. Note that due to the higher cooling rate ($\Gamma_{\mathrm{opt}} \propto Q^2 \approx 1-2$ kHz), and the relatively high background noise, the mechanical motion in the sidebands can only be observed down to a pressure of 10$^{-4}$ mbar even though we can clearly image the scattered light from the trapped particle on a CCD camera at the lowest pressures. At such a high $\textrm{N}$, no quadratic modulation is seen in the steady state data or the simulations. From the measured cooling rate, and from simulations which agree with the steady state experimental spectra at the higher pressures, we infer minimum temperatures of $\sim 1-3$mK at the lowest pressures.  Additional confidence in the derived temperatures at the lowest pressure come from the variation in the amplitudes of the power spectra that scale with the $\sqrt{\gamma_m}$ for both the simulations and the experimental data. 

A novel and useful effect of the non-trivial interaction between the Paul trap and optical field is that the optical field introduces a significant ``cavity shift'' of the Paul trap secular frequency $\omega_S$. This is somewhat analogous to the way that a DC electric field shifts the secular frequencies in a conventional Paul trap.  This shift exposes the two-way interactions in the hybrid trap where the Paul trap drives the cavity cooling while the cavity potential shifts the Paul trap frequencies.  The optically shifted secular frequencies are given by \cite{SuppInfo}:
\begin{equation}
\omega_s \simeq \frac{\omega_d}{2} \sqrt{\frac{16 \hbar A \textrm{n}}{mw^2\omega_d^2 } + \frac{8Q^2V_0^2}{(m\omega_d^2r_0^2)^2}}.
\label{Offset}
\end{equation}
Since the secular frequency $\omega_s$ is not significantly damped by the cavity cooling process, it always measurable in the cavity heterodyne spectrum, regardless of temperature. If $\omega_M$ is known then both $\textrm{n}$ and $Q$ can be extracted from the data.  This process is shown in fig.~\ref{Fig4}(b) where plots of experimentally determined secular and mechanical frequencies are shown. Also shown are the theoretical values based on cavity photon number and charge. All three values of $Q$ map well onto both the secular and mechanical frequencies. The significant shift in the secular frequencies with photon number/mechanical frequency can also be observed, demonstrating a straightforward method for calibrating our experimental data with the theoretical simulations.  

We have demonstrated cooling of levitated nanospheres from room temperatures to millikelvin temperatures by cavity cooling in a hybrid electro-optical trap.
 Cooling rates in the kHz range were measured with factor of $50$ increases possible, with a higher
 finesse cavity or by  increasing the mechanical frequency with higher photon numbers. 
This would already allow cooling to the ground state at pressures of $\sim 10^{-6}$ mBar, in the sideband resolved regime.
 By studying the cooling rate and spectral changes to the heterodyne spectra we have shown the spectral features
 of both linear $G_1$ and quadratic $G_2$ coupling. Additionally, we have shown how cooling 
rates can be enhanced by operating away from the Paul trap center where the particle is
 drawn away from the center of the antinode of the optical potential and linear cooling rates are significantly enhanced.    

Although quadratic couplings in principle offer the prospect of QND measurement of energy,
QND phonon measurement is extremely challenging given the current modest one
photon couplings of $g_1 \sim 1$ Hz and $g_2 \sim 10^{-5}$\,Hz. The advantage of levitated systems like the one demonstrated here is that they offer a very high-Q and 
long coherence times. This offers the potential for realization of the scheme for detection of quantum phonon noise in \cite{Clerk10}. There is 
a classical Paul trap drive potentially of large amplitude, which (e.g. in Fig.\ref{Fig4} ) corresponds to
a mean phonon occupancy of ${\overline n} \sim 10^8$, even if the thermal value $n_p \sim 1$ is near ground state occupancies.
Although $\omega_M/\omega_d = 8-25$ in these experiments, in principle a resonant classical drive at the mechanical frequency is possible and this will be further explored.




\begin{thebibliography}{99}


 \bibitem{AMKreview} M. Aspelmeyer, T.J. Kippenberg, F. Marquardt, Rev. Mod. Phys. {\bf 86}, 1391 (2014).

\bibitem{Lehnert11}
Teufel, J. D., Donner, T., Li, D., Harlow, J. H., Allman, M. S., Cicak, K., Sirois, A. J., Whittaker, J. D., Lehnert, K. W. and Simmonds R. W.,
Nature {\bf 475}, 359 (2011).


\bibitem{Painter11_2}
Chan, J., Mayer Alegre, T. P., Safavi-Naeini, A. H., Hill, J. T., Krause, A., Groeblacher, S., Aspelmeyer, M. and Painter, O.,
Nature {\bf 478}, 89 (2011).



\bibitem{OMIT} S. Weis et al, Science 330, 1520-1523 (2010).

\bibitem{Painter11}
Safavi-Naeini, A. H. and Painter, O.,
New J. Phys. {\bf 13}, 69 (2011).

\bibitem{Cleland13}
Bochmann, J., Vainsencher, A., Awschalom, D. D. and Cleland, A. N.,
Nature Phys. {\bf 9}, 712 (2013).

\bibitem{Lehnert14}
Andrews, R. W., Peterson, R. W., Purdy, R. W., Cicak, K., Simmonds, R. W., Regal, C. A. and Lehnert, K. W., 
Nature Phys. {\bf 10}, 321 (2014).

\bibitem{Painter13}
Safavi-Naeini, A. H., Gröblacher, S., Hill, J. T., Chan, J., Aspelmeyer, M. and Painter, O.,
Nature {\bf 500}, 185 (2013).

\bibitem{Thompson2008} J. D. Thompson, B. M. Zwickl, A. M. Jayich, Florian Marquardt, S. M. Girvin and J. G. E. Harris.
Nature 452, 72-75 (6 March 2008)

\bibitem{Clerk10} A.A. Clerk, Florian Marquardt and J.G.E. Harris, Phys. Rev. Lett. {\bf 104} 213603 (2010).



\bibitem{Nonlintheory1} K. Borkje, A. Nunnenkamp, J. D. Teufel, and S. M. Girvin Phys. Rev. Lett. 111, 053603 (2013).

\bibitem{Nonlintheory2} Andreas Kronwald  and Florian Marquardt, PRL 111, 133601 (2013).

\bibitem{jacobs09} Jacobs, K., Tian, L. and Finn, J., Phys. Rev. Lett. {\bf 102}, 057208 (2009).

\bibitem{Doolin14} C. Doolin, B. D. Hauer, P. H. Kim, A. J. R. MacDonald, H. Ramp, and J. P. Davis,
 Phys. Rev. A {\bf 89} 053838 (2014).

\bibitem{Harris15} D. Lee, M. Underwood, D. Mason, A.B. Shkarin, S.W. Hoch and J.G.E Harris,
Nature Comm. {\bf 6}, 6232 (2015).


\bibitem{PRL2015} Millen J, P. Z. G. Fonseca, T. Mavrogordatos, T. S. Monteiro, and P. F. Barker.
{ \it Phys. Rev. Lett. } {\bf 114}, 123602 (2015) 


\bibitem{Gieseler13}
Gieseler, J., Novotny, L. and Quidant, R.,
Nature Phys. {\bf 9}, 806 (2013).










\bibitem{Oreos}
Romero-Isart, O., Juan, M. L., Quidant, R. and Cirac, J. I.,
New J. Phys. {\bf 12}, 033015 (2010).

\bibitem{Chang10}
Chang, D. E. Regal, C. A., Papp, S. B., Wilson, D. J., Ye, J., Painter, O., Kimble, H. J. and Zoller, P.,
Proc. Natl Acad. Sci. USA {\bf 107}, 1005 (2010).

\bibitem{Barker10}
Barker, P. F. and Shneider, M. N.,
Phys. Rev. A {\bf 81}, 023826 (2010).



\bibitem{Pender12}
Pender, G. A. T., Barker, P. F., Marquardt, F., Millen, J. and Monteiro, T. S.,
Phys. Rev. A {\bf 85}, 021802 (2012).


\bibitem{Monteiro13}
Monteiro, T. S., Millen, J., Pender, G. A. T., Marquardt, F., Chang, D. and Barker, P. F.,
New J. Phys. {\bf 15}, 015001 (2013).

\bibitem{Li11} Li, T., Kheifets, S. and Raizen, M. G.,
Nature Phys. {\bf 7}, 527 (2011).

\bibitem{Gieseler12}
Gieseler, J., Deutsch, B., Quidant, R. and Novotny, L.,
Phys. Rev. Lett. {\bf 109}, 103603 (2012).


\bibitem{Asenbaum13}
Asenbaum, P., Kuhn, S., Nimmrichter, S., Sezer, U., and Arndt, M.,
Nature Commum. {\bf 4}, 2743 (2013).

\bibitem{Kiesel13}
Kiesel, N., Blaser, F., Deli\'c, U., Grass, D., Kaltenbaek, R. and Aspelmeyer, M.,
Proc. Natl Acad. Sci. USA {\bf 110}, 14180 (2013).


%
\bibitem{Millen14} J. Millen,	T. Deesuwan,	P. Barker	and
 J. Anders Nature Nanotechnology 9, 425–429 (2014).
















\bibitem{SuppInfo}
See Supplementary Information associated with the manuscript.



\end{thebibliography}
\end{document}